\documentclass[nofootinbib,aps,prd,twocolumn,showpacs,preprintnumbers,amsmath,amssymb]{revtex4}
\usepackage{epsfig}
\begin{document}
\title{ The strongly interacting Quark Gluon Plasma, \\ and the 
critical behaviour of QCD at imaginary $\mu$ }
\author{Massimo D'Elia $^1$}
\author{Francesco Di Renzo $^2$}
\author{Maria Paola Lombardo $^3$}
\affiliation{$^1$Dipartimento di Fisica dell'Universit\`a 
di Genova and INFN, I-16146, Genova, Italy\\
$^2$Dipartimento di Fisica dell'Universit\`a 
di Parma and INFN, I-43100, Parma, Italy \\
$^3$ INFN-Laboratori Nazionali di Frascati, 
I-00044, Frascati(RM), Italy}
       
\begin{abstract}

We explore the highly non-perturbative hot region of the QCD phase
diagram close to $T_c$
by use of an imaginary chemical potential $\mu$ 
which avoids the sign problem.  The
number density and the quark number susceptibility are consistent with
a critical behaviour associated with the  transition
line in the negative $\mu^2$ half-plane. We compare the analytic continuation
of these results with various phenomenological  models, none of which 
provides a satisfactory description of data, a failure on which we make some comments. 
These results complement and extend the information obtained 
via the  analysis of the susceptibilities evaluated at zero $\mu$, 
yielding a simple description of the candidate strongly interacting QGP phase. 
As  a byproduct of our analysis we investigate the Polyakov loop and 
its hermitian conjugate. Our data offer a vivid evidence of the importance of 
the complex nature of the functional integral measure, which results in $L (\mu) \ne \bar L (\mu)$ for a real chemical potential $\mu$.

\end{abstract}

\maketitle

\section{Introduction}

Theoretical arguments and  experimental evidence 
suggest that hadronic matter undergoes a transition to a plasma of quarks
and gluons at high temperature~\cite{Stephanov:2007fk}.
 At extremely high temperatures quarks and
gluons are nearly free, and should be described by the  Stefan-Boltzmann
law with the appropriate degrees of freedom. When temperature is not much
larger  than the critical temperature -- say, $T_c < T < \simeq 2 T_c$ --
strong interactions among the constituents give rise to 
non--perturbative effects. In short, at large T the QGP
is a gas of nearly free quarks, 
which becomes strongly interacting at lower temperatures 
$T=(1-3)\,T_c$~\cite{Shuryak:2007qs,Blaizot:2007sw}.

Several proposals have been made to characterise the properties of
the system in such non-perturbative phase. For instance the above mentioned
strong interactions might be enough to preserve bound states above $T_c$,
while coloured states might appear, deeply
affecting the thermodynamics of the system~\cite{Shuryak:2004tx}.
 Analytic techniques are being
refined more and more, so to be able to capture the features of the system
closer and closer to $T_c$~\cite{Ipp:2006ij}. 
Model theories of quasi particle physics have been considered
as well~\cite{Bluhm:2004xn,Bluhm:2006av}.
In this work we study this interesting dynamical
region by lattice QCD simulations at $T \simeq 1.1\, T_c$
(the reason for this choice will be clear in the following),
and  a nonzero baryon density.

In principle, lattice QCD simulations at non-zero baryon density
are plagued by the sign problem~\cite{Splittorff:2007ck}. However, 
it has been realised  that this problem can be circumvented
thanks to physical fluctuations, which grow relatively large in
the Quark Gluon Plasma phase. In this work we adopted the imaginary
chemical potential approach~\cite{mpl,hart1,deForcrand:2002ci,D'Elia:2002gd,
Azcoiti:2005tv,Chen:2004tb,Wu:2006su,Cea:2006yd},
which avoids the sign problem and makes
it possible conventional Lattice QCD simulations. The interested reader
might want to consult refs.~\cite{Schmidt:2006us, Philipsen:2005mj} 
for recent reviews and 
\cite{Lombardo:2004uy,Muroya:2003qs} for more pedagogical introductions
into the subject. 
\begin{figure}
\begin{center}
\epsfig{file=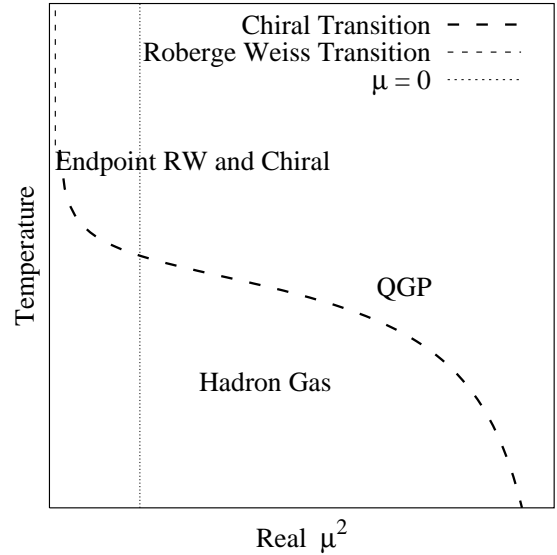,width=11 truecm}
\caption{Schematic phase diagram for four flavor QCD in the $T, \mu^2$
plane. The candidate sQGP phase is bound by the chiral (pseudo)critical
line in the negative $\mu^2$ half-plane}
\vskip -0.8 truecm
\label{fig:phaseDIAGR}
\end{center}
\end{figure}

Other studies in the quark gluon plasma phase have addressed
the higher temperature regime~\cite{Fodor:2002km,D'Elia:2004at,D'Elia:2005qu,Ejiri:2005wq, 
Kratochvila:2006jx}.  
Here we analyze in detail the non-perturbative behaviour close
to $T_c$, in the candidate strongly coupled Quark Gluon Plasma (sQGP)
region (some preliminary results have appeared in~\cite{Lombardo:2006yc}). 
We note that in the sQGP region the chiral critical line
lies in the imaginary chemical potential plane, and that such a 
chiral line ends in the proximity of the endpoint of the 
Roberge--Weiss line~\cite{Roberge}.
We focus our analysis on the particle number and its susceptibility, 
on  the chiral condensate,  and on the Polyakov loop,
and we find that the results are consistent 
with those expected of a critical behaviour associated with the critical
line at imaginary chemical potential.
Hence, the numerical results are compatible with  simple power law behaviour 
of the equation of state as a function of the imaginary chemical 
potential $\mu_I$, yielding a modified form of the Stefan-Boltzmann law.

The rest of this paper is organised as follows. Section II is devoted to
the analysis of the particle number and the chiral condensate, which are 
related by the Maxwell equation. In Section III we discuss the behaviour of
the Polyakov loop. It turns out that our approach offers a particularly
simple description of an apparent puzzle, and, at the same time, gives
a direct evidence of the phase of the determinant at nonzero, real
chemical potential. The implication on the equation of state are summarised
in Section III, while Section IV discusses our results in the light of
phenomenological proposals, and alternative lattice approaches. Last Section
is a short summing up.

\section{Thermodynamics of the Hot Phase close to $T_c$}
Let us remind ourselves of the critical lines in the phase diagram in the T, 
$\mu^2$ plane (Figure~\ref{fig:phaseDIAGR}): at high temperature there is the 
Roberge Weiss transition at $\mu = \pi T/3$, 
associated with the phase of the Polyakov Loop, ending at $T \simeq T_{RW}$.
At lower temperatures the QGP region  is limited by a chiral transition
at negative $\mu^2$, which continues into the physical chiral transition
at positive $\mu^2$, i.e. real chemical potential~\cite{deForcrand:2002ci,D'Elia:2002gd}.

While $\mu$ approaches $\pi T/3$ at a constant temperature 
$T \simeq T_{RW}$ the chiral transition
approaches the Roberg Weiss transition. Within the current numerical accuracy
the endpoint of the two transitions cannot be resolved, and the nature
of the critical behaviour around $T = T_{RW}, \mu = \pi T/3$ is 
an interesting question in itself. If T is slightly larger than $T_{RW}$
we are approaching the Roberge Weiss transition, if slightly lower we hit 
the chiral transition, and at $T = T_{RW}$ we might expect interesting
critical phenomena whose universality class is not known a priori.
Note that in the chiral limit the
$\mu=0$ transition should be of first order, and we do not expect
any tricritical point along the critical line at a real chemical potential. 
A possible occurrence of
an endpoint at finite mass in the $T, \mu$ plane depends on dynamical
details, which are not known, and are not relevant for the present study.

We have then carried out simulations  on a $16^3 \times 4$ lattice and four 
flavor of staggered fermions at $\beta = 5.1$, which, according to
our previous results, yields $T \simeq T_{RW}$,
endpoint of the RW transition,. Fermions are fully degenerate, 
with a 
bare dimensionless mass ($a$ being the lattice spacing) 
$\hat{m} \equiv m a = 0.05$. 
For our lattice the value of the (dimensionless) $\hat{\mu} \equiv \mu a$
which is relevant for the Roberg Weiss transition reads $\hat{\mu} = \pi/12$
(the temperature being $T = 1/(a N_t)$ and in our case $N_t = 4$). 
With a slight abuse we will omit in the following the 
hat-notation, nevertheless 
measuring $\mu$ and $T$ in unit of inverse lattice spacing. 

First, we check our data for the particle number 
against a simple free field behaviour. We have  numerically computed
the free field results for real chemical potential
on a $16^3 \times 4$ lattice, and $m_q = 0.05$, and we have
fitted them to an expression motivated by one dimensional QCD~\cite{Bilic:1988rw},
which turns out to be an excellent parametrisation:
\begin{equation}
n(\mu)_{free} = \frac { 3 \sinh ( \mu / T) }{ K + \cosh (\mu / T)} .
\label{eq:nfree}
\end{equation}
yielding the free field results for the number density as a function
of imaginary chemical potential
\begin{equation}
n(\mu_I)_{free} = \frac { 3 \sin( \mu / T) }{ K + \cos(\mu / T)} .
\label{eq:nfreecont}
\end{equation}

\begin{figure}
\epsfig{file=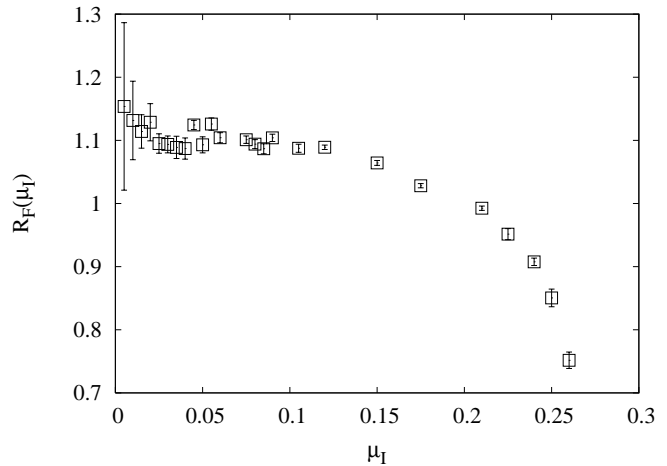,width= 9 truecm}
\caption{$R_F (\mu_I) = n(\mu_I)/n(\mu_I)_{free}$ as a function of $\mu_I$ ,
showing a very clear evidence of a deviation from a free field behaviour}
\label{fig:RATIO2free}
\end{figure}

We then considered the ratio between the numerical results and such free field
results $R_F (\mu_I) = n(\mu_I)/n(\mu_I)_{free}$ (Figure~\ref{fig:RATIO2free}).
We observe a clear dependence of $R_F (\mu_I)$ on $\mu_I$: the
results are qualitatively different from a free field  
and the discrepancy cannot be accounted
for by any simple renormalisation of the degrees of freedom.
This behaviour should be contrasted with that of Fig. 12 of 
Ref.~\cite{D'Elia:2004at} where the results at
high temperature did differ from a free field behaviour by a constant factor
very close to one.

As a second attempt at interpreting our data in terms of the simplest
parameterisations, we consider an analogy with the Hadron Resonance Gas
model~\cite{Toublan:2004ks,Karsch:2003zq}. 
We will discuss more fully these aspects in the last Section.

In a nutshell, the HRG describes the system as a gas of weakly 
interacting resonances. The pressure of the HRG model reads:
\begin{eqnarray} 
{{P (T, \mu) - P(T, 0)}\over {T^4}}&\simeq& F(T)(\cosh ({{N_c \mu_q}\over T})-1) \\
F(T)&\simeq &\int dm\rho (m) ({m\over T})^2 K_2({m\over T})
\end{eqnarray}
and the argument of the hyperbolic cosine,  $N_c \mu_q$, 
tells us that in the hadronic phase
one can only excite baryonic degrees of freedom.

Let us remind ourselves here that general arguments guarantee that  the 
partition function is periodic at imaginary $\mu$,
and that the strong coupling analysis shows that periodicity is smooth
at low temperature. Hence,  the number density reads~\cite{D'Elia:2002gd}
\begin{equation}
n(\mu_I)  = \sum_k bo_F \sin( k N_c N_t \mu_I) 
\end{equation}
When HRG holds true, one term in the Fourier 
series should suffice.
\begin{equation}
n(\mu_I)  = \frac {\partial P(\mu)}{\partial \mu} = K(T) \sin(N_c N_t \mu_I)
\end{equation}
A cross check with the Hadron Resonance which uses  the Taylor approach 
requires the computation of an infinite number of derivatives~\cite{Karsch:2003zq}, 
while the Fourier analysis -- possible with the imaginary
chemical potential approach -- needs only one parameter fit. 

\begin{figure}
\epsfig{file=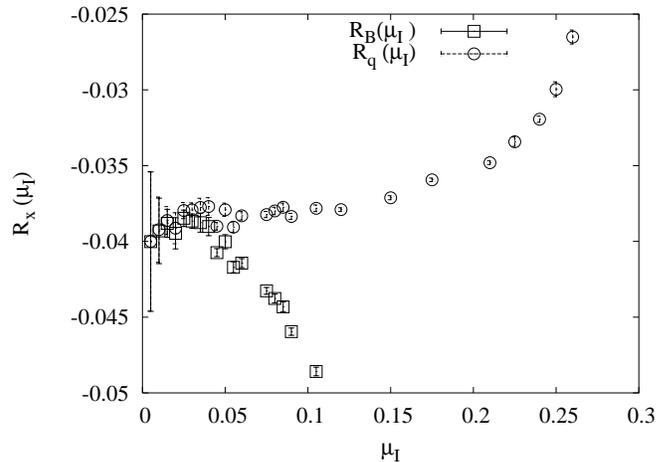,width= 9 truecm}
\caption{$R_x(\mu_I)$ (defined in eqs. 2 and 3) as a function of $\mu_I$ , 
showing the limitations of a linear approximations, as well
as the deviations from the simplest trigonometric parameterisations
motivated by a Hadron Resonance Gas model}
\label{fig:Rratios}
\end{figure}

In Figure~\ref{fig:Rratios} we display the ratios 
\begin{eqnarray}
R_B( \mu_I) &=& \frac{n(\mu_I)}{\sin (3 \mu_I/T)} \\
R_q( \mu_I) &=& \frac{n(\mu_I)}{3 \sin (\mu_I/T)} 
\end{eqnarray}

$R_B(\mu_I)$ should be a constant for a simple hadron
gas (cfr. again Ref.~\cite{D'Elia:2004at}), while, mutatis mutandis,
$R_q( \mu_I)$ should be a constant for a ``hadron gas'' made of quarks.
Both $R_B$ and $R_q$ stay constant only for a short interval of chemical
potential, indicating the region where $n(\mu_I) \propto \mu_I$. 
The deviation from a linear behaviour (for $\mu_I > 0.05$) , as well
as from the simplest trigonometric parameterisations (for $\mu_I > 0.15$),
are evident from the plot. 

We now move on and propose to describe the 
particle number in the critical region as 
\begin{equation}\label{eq:critFIT}
 n(\mu_I) = A \mu_I ({\mu_I^c}^2 - \mu_I^2)^\alpha \,.
\end{equation}
This ansatz for  $n(\mu_I)$ takes into
account that $n(\mu)$ should be an odd function of $\mu$.
Moreover, it reproduces a singular behaviour of the quark number 
susceptibility at a genuine critical point. 
Namely, the most divergent
part of the quark number susceptibility $\chi_q$ behaves as 
\begin{equation}\label{eq:chi}
\chi_q (\mu_I) \propto \frac{1}{({\mu_I^c}^2 - \mu_I^2)^\gamma}
\end{equation}

where $\gamma = 1 - \alpha $, while $\mu_I^c = \pi / 12$ if the 
critical point coincides with the Roberge-Weiss line. 

\begin{figure}
\epsfig{file=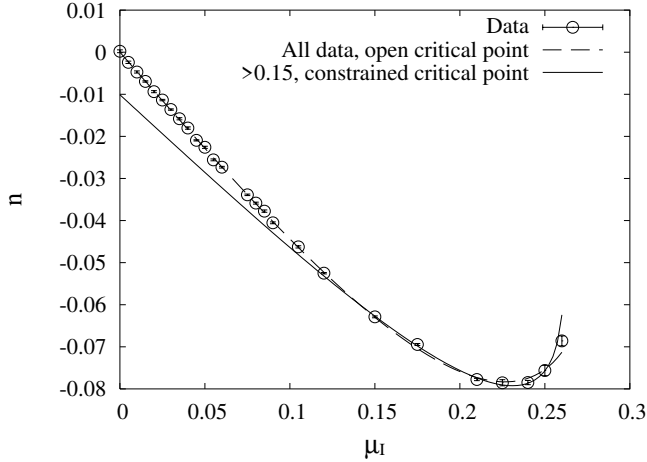,width= 9 truecm}
\caption{$n(\mu)$ fitted to the form predicted by the simple critical
behaviour at imaginary $\mu$ Eq.~(\ref{eq:critFIT})}
\label{fig:nMU}
\end{figure}

We then fit our data to Eq.~(\ref{eq:critFIT})
with $\mu_I^c$ either open or constrained. 
A fit to our entire interval with unconstrained $\mu_I^c$ gives 
$A = -0.94(4)$, $ {\mu_I^c}^2 = 0.0804(2)$,  $\alpha = 0.28(2)$
with a reduced $\tilde\chi^2 =   2.4$. We checked the stability of these
results by choosing different ranges in chemical potential, and
we obtained the exponent $\alpha$ ranging from 0.34(8) and 0.26(3),
$ {\mu_I^c}^2$ between 0.078(4) and 0.091(12), with reduced $\tilde\chi^2 $
ranging between 1.8 and 5. (see Figure~\ref{fig:nMU} for two representative fits).

If we constraint 
$ {\mu_I^c}^2 = (\pi/12) ^2 $ the quality of the fits decreases giving a reduced 
$\tilde\chi^2 \simeq 12 $.  If we limit the fitting interval
to $\mu_I > 0.15$, we need to add a constant to the function to approximate
the regular component in this interval to obtain a reduced $\tilde\chi^2 = 3.5$,
with $\alpha = 0.12(1)$. All in all, constraining $\mu_I^c$ does not
improve the results.

\begin{figure}
\epsfig{file=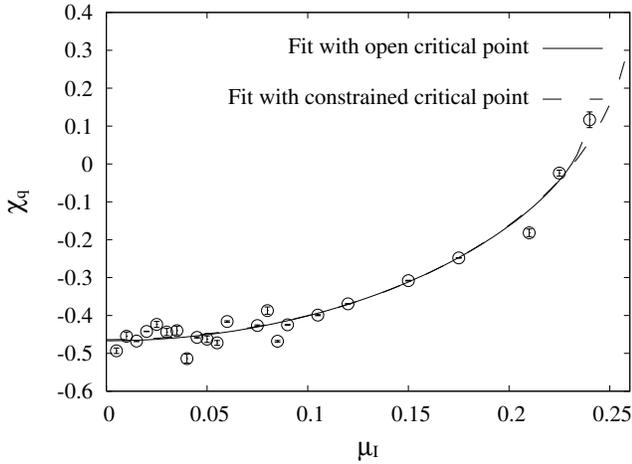,width= 9 truecm}
\caption{The quark number susceptibility obtained by numerical
differentiation of the results for the quark number}
\label{fig:chiQ}
\end{figure}

We now go back to the quark number susceptibility as entailed 
in Eq.~(\ref{eq:chi}), for which our results 
indicate $\gamma = 1 - \alpha \simeq 0.7$. 
Figure~\ref{fig:chiQ} shows $\chi_q$  obtained
by numerical differentiation of $n(\mu_I)$. The numerical quality
is of course poor, but, anyway, a fit to the form of Eq.~(\ref{eq:chi}) 
with an open $\mu_I^c$ gives  $\gamma = 0.66(16)$,
while a fit with constrained $\mu_I^c$ gives $\gamma = 0.44(22)$,
in agreement with the above estimates, within the large errors.

The chiral condensate $<\bar \psi \psi>$ can be inferred from
the chiral equation of state  in either phases, and also in the presence
on an explicit chiral symmetry breaking term. 
To make a closer contact with thermodynamics, we  consider 
the Maxwell relation (the temperature is constant and its dependence is
omitted)~\cite{Kogut:1983ia}:
\begin{equation}
\frac{\partial n(\mu, m)} {\partial m } = 
\frac{\partial < \bar \psi \psi> (\mu,m) } {\partial \mu}.
\label{eq:max}
\end{equation}
Considering the $m$ dependence in the expression for $n(\mu,T)$
\begin{equation}
n(\mu, m) = F(m) \mu ({\mu_I^c}^2(m) - \mu^2)^\alpha
\end{equation}
where $F(m)$ depends only on the quark mass, and using Eq.~(\ref{eq:max}) 
we arrive at
\begin{eqnarray}
<\bar \psi \psi> (\mu) &=& H(m)  ({\mu_I^c}^2(m) -
\mu^2)^{\alpha + 1}   \\ & + & K(m)  ({\mu_I^c}^2(m) - \mu^2)^\alpha + 
<\bar \psi \psi>_0 \nonumber
\end{eqnarray}
where $H(m), K(m)$ depends only on the mass  and  
$<~\bar\psi\psi>_0$ is an integration constant
which can be fixed e.g. by the chiral condensate at
zero chemical potential. 

We have then fitted the chiral condensate to the leading term 
at $ \mu \simeq \mu_c$
\begin{equation}
< \bar \psi \psi> = K ( b - \mu^2)^{\alpha_c} + A 
\end{equation}
obtaining a nice fit with a reduced $\tilde\chi^2 = 0.79 , A = 0.552(6) , b =
0.06628(8), K = -0.63(2)$ and $\alpha_c = 0.47 (2)$ in reasonable agreement 
with $\alpha$ estimated from the number density. By constraining the
fitting interval $\mu_I > 0.2 $ the sub-leading contributions are less
important, and $\alpha_c = 0.32 (12)$ gets even closer to $\alpha$.

It might be interesting to compare this critical behaviour with that of 
the endpoint of QCD from model field theories~\cite{Hatta:2002sj}.

\begin{figure}
\epsfig{file=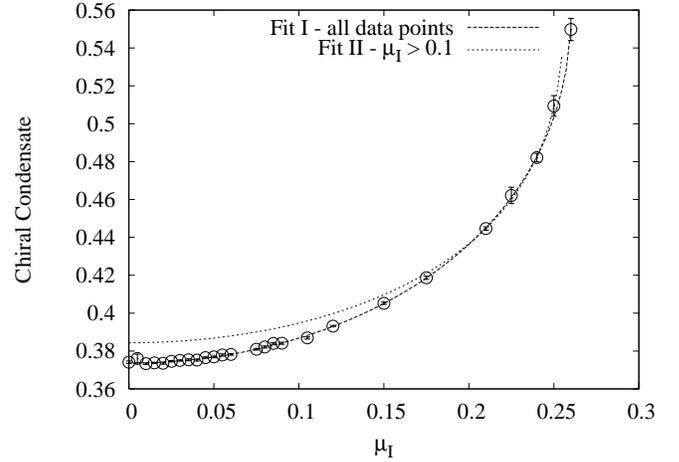,width = 9 truecm}
\caption{Numerical results for the chiral condensate, and the results of
  the fit to a functional form inferred from 
the Maxwell equation Eq.~(\ref{eq:max}).}
\label{fig:chiral}
\end{figure}

\section{The Polyakov Loop}
\begin{figure}
\epsfig{file=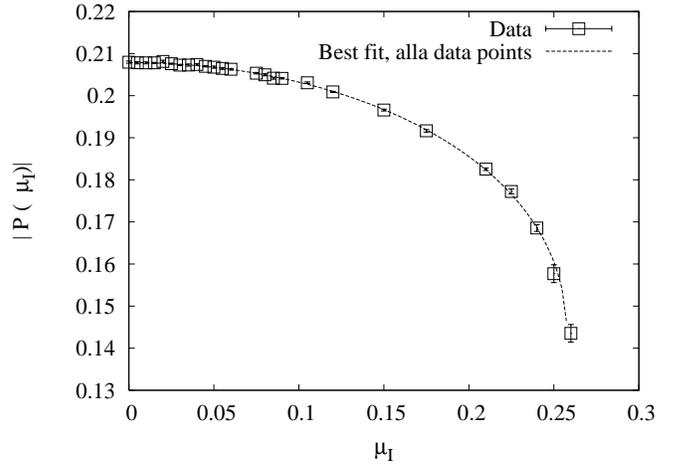,width= 9 truecm}
\caption{The Polyakov loop fitted to the form predicted by a simple
critical behaviour at imaginary $\mu$  Eq.~(\ref{eq:crip}). }
\label{fig:ABSPoly}
\end{figure}

In the same spirit we  have fitted the traced Polyakov 
loop $L = \mbox{Tr} P $ to
a power law form 
\begin{equation}
L(\mu_I) \propto ({\mu_I^c}^2 - \mu_I^2)^\beta
\label{eq:crip}
\end{equation}
Figure~\ref{fig:ABSPoly} displays the results of the fit of the absolute
value of the Polyakov loop, which looks indeed satisfactory. 

It is interesting to look in more detail into the behaviour of 
$L = \mbox{Tr} P$.

The Polyakov loop $P$ satisfies the same relation as the 
quark propagator at nonzero chemical potential~\cite{Kogut:1994eq}

\begin{equation}
P(\mu) = P^\dagger (-\mu)
\label{eq:polsym}
\end{equation}

This relation implies that, while both $L = \mbox{Tr} P $ and $\bar L = \mbox{Tr} P^\dagger $ 
are real at real chemical potential, $L \ne \bar L$, as noted
in~\cite{Karsch:1985cb, Allton:2002zi, Dumitru:2005ng, Roessner:2006xn,
Kratochvila:2006jx}. We will show that the results at imaginary chemical 
potential  offer a particularly simple illustration of these ideas,
as well as a direct evidence of the complex phase of the determinant.

The asymmetry at real chemical potential 
is easily understood by considering the distribution  of the Polyakov loops
in the complex plane: for $\mu=0$, 
since the $Z_3$ centre symmetry is broken by the 
dynamical quarks, the  root corresponding to the phase $\phi = 0$ is preferred, 
and the average is non zero. 
A non zero, positive chemical potential encourages
forward propagation: the distribution of the phases is further
peaked at $\phi=0$, while the two other phases have the same probability.
Hence,  the Polyakov loop remains real, and the 
final average is again real, different from zero, and slightly larger
than the one at zero density. 
$\bar L$  instead describes backward propagation: again 
the Polyakov loop remains real, however its length is reduced,
hence $L (\mu) \ne \bar L (\mu)$. 

Notice that at a first na\"{\i}ve look it may sounds strange that,
while configuration by configuration the Polyakov loop and its
hermitian conjugate are always the complex conjugate of each other,
their expectation values, even being real, differ from each other. However it should be
clear that the complex nature of the functional integral measure plays
an essential role in this respect, since the real part of 
the expectation value is not just the expectation value of the real
part. In that sense the fact that  $L (\mu) \ne \bar L (\mu)$
is directly linked to the complex nature of the fermion determinant,
thus also giving a qualitative feeling about the severeness of the sign problem.

An apparent puzzle then arises when one considers the behaviour at imaginary
chemical potential: there the measure is real and one can show that
the absolute value of $L$ and $\bar L$ are equal as well as their real parts.
What is then the fate of the asymmetry which is present at real
chemical potential?

Consider 
\begin{equation} 
L_{o/e} (\mu) \equiv L (\mu) \pm L(-\mu) = L (\mu) \pm \bar L(\mu)\, ,
\end{equation}
where Eq.~(\ref{eq:polsym}) has been used in last equality.
$L_{o/e}$ are respectively even and odd in $\mu$. Remember 
that the analytic continuation to imaginary chemical potential of
an even function is real, while the analytic continuation of
an odd function is purely imaginary. 
Hence, the analytic continuation of the even observable 
$ L_e(\mu) = L (\mu) + \bar L (\mu) $ at imaginary chemical
potential is the real part of $L (\mu_I)$;  while the analytic
continuation of $ L_o(\mu) = L(\mu) - \bar L (\mu)$  
is  the imaginary part of $L (\mu_I) $ at imaginary $\mu$. 
$L$ itself has no definite $\mu$-parity and 
its analytic continuation develops an imaginary part.

\begin{figure}
\epsfig{file=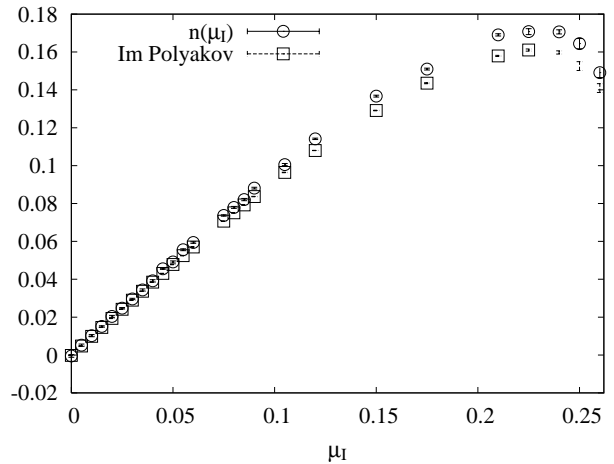,width= 9 truecm}
\caption{The imaginary part of the Polyakov loop, divided by the coefficient
of the linear term of a fit at small $\mu$,  as a function of the
imaginary chemical potential, demonstrating the relevance of the
phase of the determinant for a real chemical potential; in the same plot
we show the number density, again normalised by the coefficient of the
liner term. }
\label{fig:IMPoly}
\end{figure}

We conclude that we must search for the analytic continuation
of the asymmetry $L \neq \bar L$ which is present at real chemical 
potential in the imaginary part of the Polyakov loop, which is non-zero
in presence of an imaginary chemical potential.
Figure~\ref{fig:IMPoly} shows the imaginary part of $L(\mu_I)$: 
it is different from
zero, offering 
 a clean, direct evidence of the asymmetry $L \neq \bar L$, hence of the
complex phase of the determinant
at real chemical potential.  In the same Figure~\ref{fig:IMPoly} 
we have also plotted
$n(\mu_I)$ (both $L$ and $n$ with an appropriate normalisation),
to show its correlation with $L(\mu_I)$. This correlation is
in agreement with the lattice interpretation of the number operator $n$
as 'counting' the links winding forwards minus those winding backwards,
and it should become exact in the heavy quark limit when the model
reduces to a Polyakov loop model~\cite{Blum:1995cb,DePietri:2007ak}.

Let us now come back to the critical behaviour of $|L|$: 
since Im$(L(\mu_I) \propto n (\mu_I)$, we might expect that $|L (\mu_I)| $ 
approaches zero with a similar power law.
The fit of Figure~\ref{fig:ABSPoly} gives an exponent of 0.45(2), in the same range as 
$\gamma$. It would be interesting, of course, the study of the string tension
and other correlators in the  same range of chemical potentials.

\section{Analytic continuation to real chemical potential}
\begin{figure}
\hskip -0.5 truecm \epsfig{file=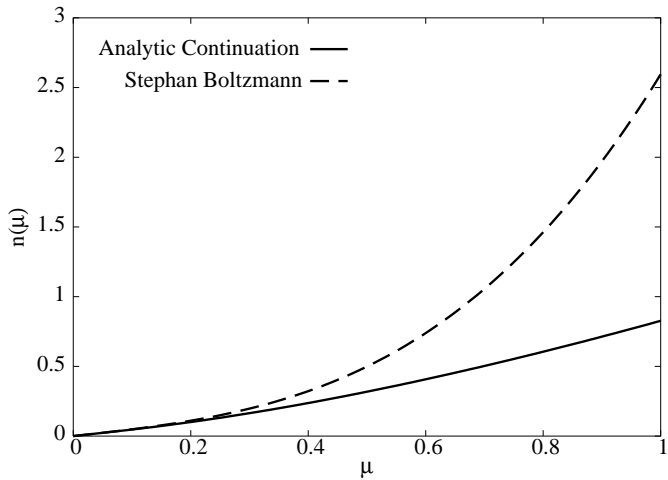,width= 9 truecm}
\caption{$n(\mu)$ from analytic continuation, together with 
a free field behaviour is shown for comparison. The fits suggest that the
slower increase observed in the interacting case with respect to the free case
can be described by on overall exponent smaller than one}
\label{fig:CONTnmu}
\end{figure}

Finally, we can  analytically continue\footnote{The singularity in the complex $\mu$ plane might well limit
the radius of convergence of the Taylor expansion. However, the
proposed -- Pade' like'-- parametrisation does not suffer from this
problem~\cite{Lombardo:2005ks,Cea:2006yd}  
and  it should not come as a surprise that we are able to analytically
continue beyond the radius of convergence of the Taylor series: the
analytic continuation is then valid for all the real values of the chemical
potential $\mu$.} our results for  n($\mu_I$), obtaining 
\begin{equation}
 n(\mu) = A \, \mu \, ({\mu_I^c}^2 + \mu^2)^{\alpha}
\label{eq:cont}
\end{equation}
with $\alpha \simeq .3$. The results are shown in Figure~\ref{fig:CONTnmu}.
In the same diagram we also plot the free field results, as an indicative
comparison. 
Note that  coefficient A has a non trivial dimension, again
indicating that the system is not free.

It is then amusing to notice that by using  simple 
arguments from the theory of critical phenomena we arrive at a modified
(lattice) Stefan-Boltzmann law, which would correspond to $ \alpha = 1$, 
and a large ${\mu_I^c}^2 \simeq 0.5$. In this framework a large
$\mu_I^c$  can be interpreted
as a spinodal point at imaginary chemical far away from the Roberge Weiss line.

Obviously, Eq.~(\ref{eq:cont})  accounts for a slower increase of the
particle density closer to $T_c$ than in the free case. This is
expected on physical grounds, as well as 
from the behaviour of the susceptibilities~\cite{Gottlieb:1988cq,Bernard:2002yd,Gavai:2003mf,Choe:2002mt,Gavai:2003nn,
Allton:2002zi}, and of course accounts 
for the behaviour observed in Figure~\ref{fig:RATIO2free}.

From the results above, we conclude
that the data in the candidate region for a strongly coupled QCD
are  accounted for by a conventional critical behaviour:
clearly, a free field behaviour would have been incompatible 
with it.
In other words, the nonperturbative features of the plasma are closely
related with the occurrence of the critical line at negative $\mu^2$.

\section {Thermodynamics and particle content}

Early in this note we have contrasted our data with some
very simple parameterisations  motivated by the
HRG approach. Here we discuss this point in some more detail,
in particular we wish to examine the proposal that in the deconfined, 
strongly  interacting region considered here 
one might observe either coloured  and colourless~\cite{deForcrand:2000jx}
 bound states:
in short, at large T the QGP
is a gas of nearly free quarks, 
which becomes strongly interacting at lower temperatures $T=(1-3)\,T_c$,
see e.g.~\cite{Shuryak:2006se,Blaizot:2006up} for recent reviews and
a complete set of references. 

In Ref.~\cite{Ejiri:2005wq}
the following  parametrisation was proposed for the 
contribution of the coloured states to the subtracted pressure
$\Delta P_C = P_C (T, \mu) - P_C (T, 0)$ 
(we slightly simplify the notation):
\begin{eqnarray}
\frac{ \Delta P_C}{T^4} &= & \left( F_q(T)\right)
\left( \cosh(\mu_u/T) +\cosh(\mu_d/T) \right) \nonumber \\
&&+ F_{qq}(T) ( \cosh(2 \mu_u/T) +
\cosh(2 \mu_d/T) \nonumber \\
&&+ \cosh((\mu_u+\mu_d)/T) )  \, .
\label{eq:cologas}
\end{eqnarray}
The susceptibilities at zero chemical potential 
 can be easily computed  from Eq.~(\ref{eq:cologas}),
and we recognise  that their ratios allow the identifications
of the relevant degrees of freedom. These prediction for the susceptibilities
ratio was contrasted with the numerical results, finding a poor 
agreement. 

The imaginary chemical potential approach gives the possibility to
check directly the consistence of various phenomenological models
 by analytically continuing
from real to imaginary $\mu$. 
We can subject our data to the same analysis by 
analytically continuing Eq.~\ref{eq:cologas} from real to
imaginary chemical potential. Setting 
$\mu_{isospin} = \mu_u - \mu_d =0$, and including the contribution
from  baryons and tetraquarks we get: 
\begin{eqnarray}
\frac{\Delta P}{T^4} &=&  F_q(T)\cos(\mu/T)) + F_{qq}(T) \cos(2 \mu/T)  
\\
&+& F_{qqq}(T) \cos (3 \mu /T)+ + F_{qqqq}(T) \cos (4 \mu /T)  \nonumber
 \end{eqnarray}
giving in turn:
\begin{eqnarray}
n(\mu_I,T) &=&F_q(T) \sin (\mu_I/T) + 2 F_{qq} (T) \sin (2 \mu_I/T)  
\\ & + & 
3 F_{qqq} (T) \sin (3 \mu_I /T) + 4 F_{qqq} (T) \sin (3 \mu_I /T) \nonumber
\end{eqnarray}

\begin{figure}
\epsfig{file=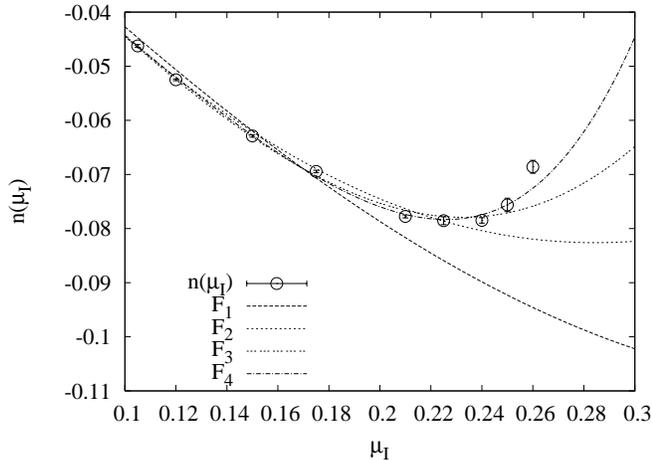,width= 9 truecm}
\caption{Results of the fits of $n(\mu_I)$ to the trigonometric functions
Eq.~(\ref{eq:FOURIERnmu}), see text for details}
\label{fig:nmuFOURIER}
\end{figure}

From the point of view of the imaginary chemical potential analysis,
checking these forms correspond to perform a Fourier analysis of
our results.We have then fitted our data to the form
\begin{equation}
F_K(\mu_I) = {\sum_j}_1^K F_j \sin (j \mu_I/T) 
\label{eq:FOURIERnmu}
\end{equation}
The results of the fits are shown in Figure~\ref{fig:nmuFOURIER}. 
The reduced $\tilde\chi^2$ ranges
from 84 to 2.85 ($F_1$ to $F_4$) but the errors on the parameters
grow big and the parameters themselves are not stable. We summarise the
results  in  Table 1, and  we conclude
that, even if the trigonometric fits might eventually converge, it is
hard to attach any simple physical interpretation of the parameters
$F_1, F_2, F_3, F_4$  as
contribution from free quarks, diquarks, baryons and tetraquarks.

\begin{table}
\caption{Parameters of the Trigonometric Fits}
\begin{tabular}{ccccc}
\hline
$F_1$ &  $F_2$ &  $F_3$ &  $F_4$ & $\chi^2/d.o.f.$ \\
\hline
 -0.110(1) &  -- &  -- & -- & 84 \\
\hline
-0.071(3) & -0.023(2) & -- & --  & 11.11 \\
\hline
0.028(15) & - 0.114(14) &  0.029(4) &  -- & 4.18 \\
\hline
0.43(11) & -0.55(13) & 0.257(66) & -0.049(14) &   2.85 \\
\hline
\end{tabular}

\end{table}

This result is not unexpected, as the Fourier parametrisation 
Eq.~(\ref{eq:FOURIERnmu}) is not compatible with the critical fits 
Eq.~(\ref{eq:critFIT}).
Does this mean that the occurrence of coloured  bound states is ruled out?

Not really: apart from the fact that 
extending an Hadron Resonance Gas 
description to coloured states is anyway a non-trivial assumption, 
given the non-trivial interactions that one could expect. 
Consider  that the masses
themselves should depend on the chemical
potential~\cite{Liao:2005pa,Bluhm:2004xn} - 
the so-called BKS effect. From the perspective of
the present study, the BKS effect is indeed very natural, 
 in view of the phase diagram in 
Figure~\ref{fig:phaseDIAGR}, and the related  analysis
of the data in term of critical behaviour which we have presented above.
Remember, in fact,  that the coefficients of the Hadron Gas parametrisation
in  Eq~(\ref{eq:cologas}) above 
represent the contribution of the spectrum of resonances, hence their
temperature dependence is inferred from the one of the masses.

For instance, in Ref.~\cite{Shuryak:2004tx} it was proposed that
\begin{equation}
M_{\rm coloured}\approx 11.5\,T_c\left((T/3T_c)^{0.5}+0.1T_c/(T-T_c)\right) 
\end{equation}
yielding the decoupling of the coloured masses at the critical point.
A similar decoupling should take place at the critical line at negative
$\mu^2$, which is not compatible with  the simple factorisation of the 
terms depending of temperature and fugacities implied by the 
HRG model Eq~(\ref{eq:cologas}).

In Ref.~\cite{Liao:2005pa} it was underscored that
if the derivatives of the masses with respect to the chemical
potential 
\begin{equation}
M '' (T) = \frac{ \partial ^ 2 M (T, \mu)}{\partial \mu ^2} (T, \mu = 0)
\end{equation}
are large enough, the simple interpretation of the zero chemical potential 
susceptibilities  as probes of particle contents has to be revised,
and, more generally, the decoupling of the prefactor and the simple
trigonometric factors predicted by the HRG model is no longer true. Hence, 
we cannot use simple forms as those of 
 Eq~(\ref{eq:cologas}) to assess the particle content of the sQGP gas. 

Our analysis supports this point of view. 
It is interesting to note that indeed a recent direct calculation~\cite{Doring:2007uh} of the
coloured spectrum indicates the survival of heavy coloured states above $T_c$.

\section{Summary and Outlook}

We have studied the critical behaviour of the system in proximity
of the critical endpoint of the chiral and RW line in the negative
$\mu^2$. We have given a simple description of
the non-perturbative features of the sQGP phase, based on the analysis
of the critical behaviour in the imaginary $\mu$ plane.
We have proposed an EOS of the form

\[
n(\mu) = A \, \mu \, ({\mu_I}_c^2 + \mu^2)^\alpha
\]

with $\alpha \simeq 0.3$,
which accounts nicely for all the features of the numerical data. The exponent 
would read $\alpha=1$ for a Stefann--Boltzmann-like law. 

From a more mathematical point of view, the proposed parametrisation
is a Pade' approximant of order [2,1], as appropriate in the standard
application to critical phenomena. The results thus obtained can be,
in principle,   analytically continued within the entire analyticity domain.
Practical limitations - discussed at length in previous work - 
do arise because of numerical accuracy.

As for the particle content of the system, our results suggest that 
a fit to 
\begin{eqnarray}
\frac{\Delta P}{T^4} &=&  F_q(T)\cos(\mu/T)) + F_{qq}(T) \cos(2 \mu/T)  
\nonumber \\
&+& F_{qqq}(T) \cos (3 \mu /T)+ + F_{qqqq}(T) \cos (4 \mu /T)  \nonumber
 \end{eqnarray}
cannot afford any definite conclusion. This does not come as a surprise. 
The masses themselves, hence the coefficients,
will depend on $\mu$ in some complicated way, which should anyhow 
conjure to give the simple behaviour observed in the data.
The fact that the masses themselves should depend on the chemical
potential while approaching the critical endpoint 
offers a simple realisation of the BKS mechanism.
It should also pointed out that extending a Hadron Resonance Gas 
description to coloured states is anyway a non-trivial assumption, 
given the non-trivial interactions that one could expect. 

It would be very interesting to confront the numerical results in
a broader range of temperatures with these ideas, as well as
with analytic calculations and phenomenological models.
Future work should hopefully be able to give a coherent account of critical
behaviour, high temperature expansions and particle contents in the
region of the strongly interactive Quark Gluon Plasma. 
We hope that the simple description offered here
might be of help in building such a complete picture.

\section{Acknowledgements}

We wish to thank Claudia Ratti, Burkhardt K\"ampfer, Aleksi Vuorinen,
Francesco Becattini and Philippe de Forcrand for helpful conversations
and correspondence. We  acknowledge
partial support from the Galileo Galilei Institute in Florence,
under the {\em High Density QCD} program. The numerical simulations were 
carried out on the APEmille computer
installed in Milano-Bicocca and we would like to thank our colleagues 
in Milano and Parma for their kind support.

\end{document}